\documentclass[a4paper,11pt]{article}
\usepackage{pos}
\usepackage{hyperref}
\usepackage{caption}
\usepackage{subcaption}

\title{The light Roberge-Weiss tricritical endpoint at imaginary isospin chemical potential}
 \ShortTitle{The light RW tricritical endpoint at imaginary isospin}

\author[b]{Bastian B. Brandt}
\author[a]{Volodymyr Chelnokov}
\author[a]{Francesca Cuteri}
\author[b]{Gergely Endr\H{o}di}
\author*[a]{Christopher Winterowd}

\affiliation[a]{Institut für Theoretische Physik,\\
Goethe-Universität Frankfurt am Main, \\
  Max-von-Laue-Str. 1, 60438 Frankfurt am Main, Germany}

\affiliation[b]{Fakultät für Physik,\\
Universität Bielefeld, \\
D-33615 Bielefeld, Germany}

\emailAdd{brandt@physik.uni-bielefeld.de}
\emailAdd{chelnokov@itp.uni-frankfurt.de}
\emailAdd{cuteri@itp.uni-frankfurt.de}
\emailAdd{endrodi@physik.uni-bielefeld.de}
\emailAdd{winterowd@itp.uni-frankfurt.de}

\abstract{We discuss results for the Roberge Weiss (RW) phase transition at nonzero imaginary baryon and isospin chemical potentials, in the plane of temperature and quark masses. Our study focuses on the light tricritical endpoint which has already been used as a starting point for extrapolations aiming at the chiral limit at vanishing chemical potentials. In particular, we are interested in determining how imaginary isospin chemical potential shifts the tricritical mass with respect to earlier studies at zero imaginary isospin chemical potential. A positive shift might allow one to perform the chiral extrapolations from larger quark mass values, therefore making them less computationally expensive. We also present results for the dynamics of Polyakov loop clusters across the RW phase transition.}

\FullConference{%
 The 38th International Symposium on Lattice Field Theory, LATTICE2021
  26th-30th July, 2021
  Zoom/Gather@Massachusetts Institute of Technology
}

\newcommand{\beq} {\begin{eqnarray}}
\newcommand{\eeq} {\end{eqnarray}}
\newcommand{\nn}{ \nonumber} 

\newcommand{\tr}{ {\rm Tr} \, }

\begin{document}
\maketitle

\section{Introduction}
Despite recent progress, our understanding of the phase diagram of QCD is still incomplete. At nonzero, real quark chemical potential, $\mu_q\neq0$, the sign problem hinders attempts to study the theory using traditional lattice methods, with the exception of the special case of QCD at pure isospin chemical potential $\mu_I = (\mu_u-\mu_d)/2$. Thus, the current understanding of the phase diagram incorporates predictions from effective models as well as nonperturbative studies at modest values of $\mu_q$. QCD with purely imaginary quark chemical potentials, $\mu_q = i\tilde{\mu}_q$, does not suffer from the sign problem and exhibits a rich phase structure which has implications for the phase diagram at real chemical potential.

Typically one considers the case of an imaginary baryon chemical potential, $\mu_B=i\tilde{\mu}_B=i(\tilde{\mu}_u+\tilde{\mu}_d)/3$ in the $N_f=2$ case, with equal quark chemical potentials $\tilde{\mu}_u$ and $\tilde{\mu}_d$.
In this case there exist vertical first-order lines in the $T-\tilde{\mu}_q$ plane at certain critical values of the imaginary quark chemical potential, $\tilde{\mu}_{q,c}/T = (2k+1)\pi/3, k \in \mathbb{Z}$. These first-order lines terminate at an endpoint at some $T_c$. This phase structure was predicted by Roberge and Weiss \cite{ROBERGE1986734} and has been confirmed by lattice studies \cite{PhilipsenDeForcrandPRL,BonatiRW2011,BonatiRW2014}. The nature of the endpoint of these first-order lines depends on the quark masses and is a question of interest to lattice practitioners due to its implications for establishing the order of the transition in QCD at $\mu_q=0$ in the chiral limit. In particular, upon finding the tricritical mass, one could, in principle, extrapolate along the $Z_2$ critical line in the $N_f=2$, $(m,\mu_q)$-plane towards $\mu_q=0$. Previous studies using unimproved staggered and Wilson fermions have located both the light and heavy tricritical quark masses for $N_f=2$ \cite{PinkePhilipsen2014,Czaban2016,PhilipsenSciarra2020}. The RW endpont was also studied in setups with improved fermion discretizations and/or number of flavors $N_f$ \cite{Bonati2016,Wu2017,Bonati2019}. In \cite{PhilipsenSciarra2020}, cutoff effects were also quantified which helped shed light on the possible fate of the chiral first-order region as the continuum limit is approached. 

One can also extend the study of imaginary chemical potentials by considering both an imaginary baryon and isospin chemical potential $\mu_I = i\tilde{\mu}_I$. An exploration of the theory in the $\tilde{\mu}_B-\tilde{\mu}_I$ plane reveals a rich phase structure which generalizes the original RW study~\cite{AmineBachelorThesis,ChabaneEndrodi2021}. Apart from this, one is also interested in the effect that the imaginary isospin chemical potential has on the light tricritical mass. Namely, a positive shift in the tricritical point could aid attempts to determine the nature of the phase diagram in the chiral limit. This is due to the fact that chiral extrapolations could be done using simulations performed at heavier quark masses which would greatly reduce the overall cost of these calculations.

\section{Setup}

\begin{figure}
        \centering
        \includegraphics[scale=0.5, angle=0]{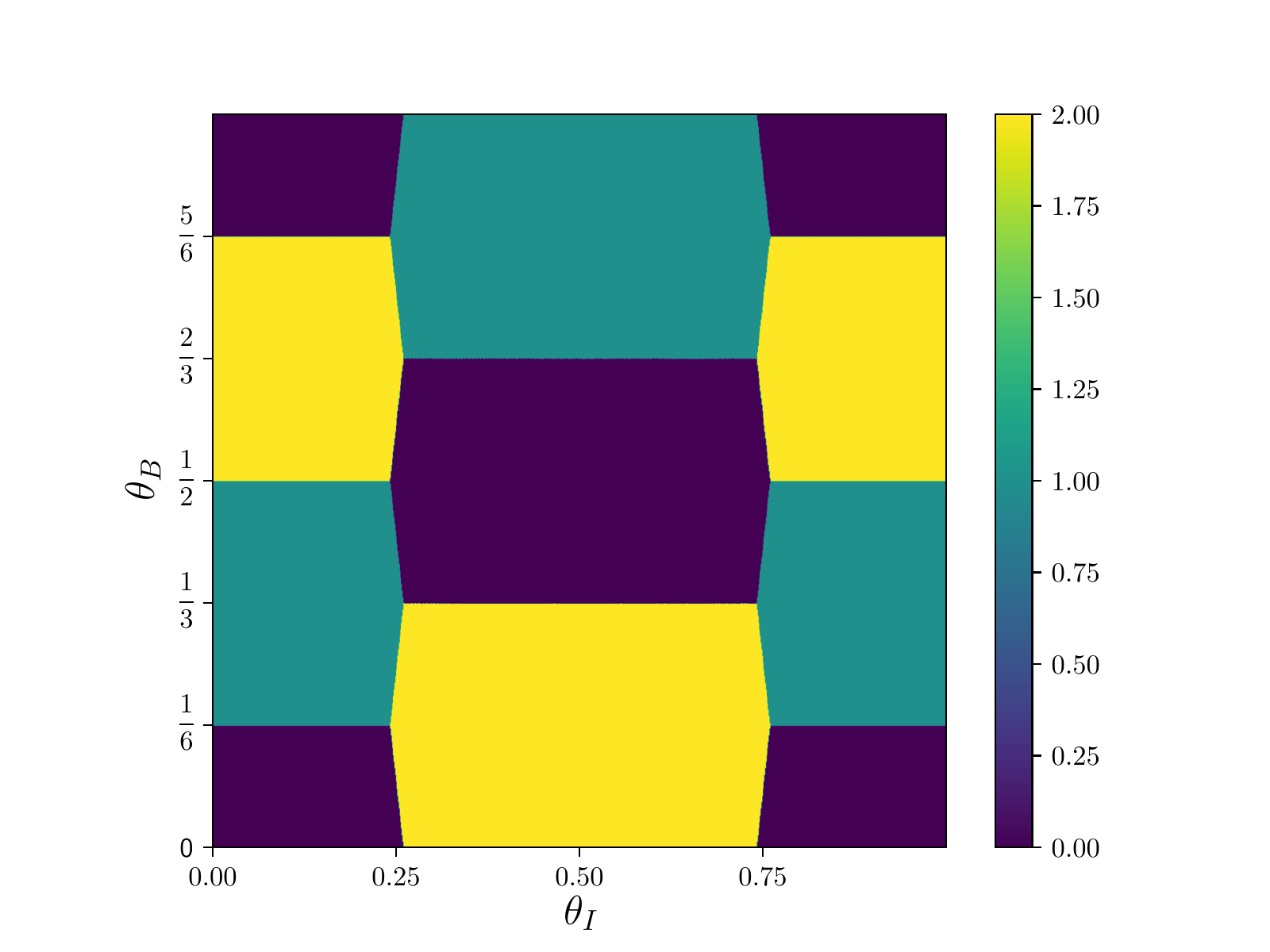}
        \caption{Phase diagram from \cite{ChabaneEndrodi2021}, displaying the structure of the theory in the $(\theta_q,\theta_I)$-plane. The colors denote the orientation of the volume-averaged Polyakov loop for which the free energy has a minimum in the one-loop approximation. This picture holds at $T\to \infty$ while the situation close to the RW critical point remains an open question.}
        \label{fig:carpet_plot}
\end{figure}

We use $N_f=2$ mass-degenerate flavors of rooted, unimproved staggered fermions which are described by the following partition function
\beq \label{eq:PartitionFunction}
Z(T, \mu_1, \mu_2) = \int \mathcal{D} U e^{-S_G[U]} \left( \det M(\mu_1) \right)^{1/4} \left( \det M(\mu_2) \right)^{1/4}, 
\eeq 
where $S_G$ is the standard Wilson plaquette action and the unimproved staggered fermion operator at nonzero imaginary chemical potential reads
\beq \nn
M(\mu_i)_{i,j} = am \delta_{i,j} + \frac{1}{2} \sum^3_{\nu=1} \eta_{i,\nu}\left( U_{\nu}(i) \delta_{i,j-\hat{\nu}} -U^{\dagger}_{\nu}(i-\hat{\nu}) \delta_{i,j+\hat{\nu}} \right) \\+  \frac{1}{2}\eta_{i,0}\left( e^{ia\tilde{\mu}_i} U_{0}(i) \delta_{i,j-\hat{0}} -e^{-ia\tilde{\mu}_i}U^{\dagger}_{0}(i-\hat{0}) \delta_{i,j+\hat{0}} \right).
\eeq 
It is convenient to write the chemical potential dependence of the theory in terms of the variables $\theta_{u,d} \equiv \tilde{\mu}_{u,d}/T =  aN_{\tau}\tilde{\mu}_{u,d}$, which can be expressed as $\theta_u =  \theta_q + \theta_I$ and $\theta_d = \theta_q - \theta_I$. One can show that (\ref{eq:PartitionFunction}) is periodic in both $\theta_q$ and $\theta_I$ with period $2\pi$, and is also invariant under the transformations $(\theta_1, \theta_2)\to (\theta_2, \theta_1)$, $(\theta_1, \theta_2)\to (-\theta_2, -\theta_1)$, and $(\theta_q, \theta_I)\to(\theta_q+\pi, \theta_I+\pi)$. The usual RW transformation, which consists of $\theta_q \to \theta_q +2\pi k/3,~k \in \mathbb{Z}$ and a gauge transformation, $G(x) \in SU(3)$ satisfying $G(\vec{x}, \tau+N_{\tau}) = H G(\vec{x}, \tau)$, with $H \in Z(3)$, also applies. These relations greatly constrain the phase diagram of the theory in the $(\theta_q,\theta_I)$-plane. Initial perturbative studies, confirmed by recent lattice calculations, suggest a ``carpet''-like phase structure describing the sector of the volume-averaged Polyakov loop which corresponds to the minimum of the free energy. This can be seen in fig.~(\ref{fig:carpet_plot}).  

Our simulations used a bare quark mass of $am \in \{0.04,0.05,0.06\}$ and a fixed temporal extent of $N_{\tau}=4$. These masses were chosen taking into account the result for the light tricritical mass, $m_{tc} = 0.043(5)$ at $N_{\tau}=4$ in \cite{BonatiRW2011}. The gauge configurations were generated with the RHMC algorithm. Our calculations employed an aspect ratio, $N_s/N_{\tau} \in [3,6]$, where the spatial volume is given by $V=N^3_s$. This is in line with previous studies of the RW tricritical point and allowed us to accurately perform a finite-size-scaling analysis of the normalized fourth moment of the imaginary part of the volume-averaged Polakov loop, $P = \frac{1}{V} \sum_{\vec{x}} \tr \prod_{\tau} U_0(\vec{x},\tau)$. We have fixed the imaginary chemical potentials such that we sit at the point $(\pi, \pi/6)$ in the $(\theta_q,\theta_I)$-plane. At this point, one sits at the critical value in $\theta_q$ which separates sectors differing in the sign of $\Im P$, and is, presumably, far enough away from the first critical value of $\theta_I$. 

\section{Results}
\begin{figure}
        \centering
        \begin{subfigure}{.45\textwidth}
  \centering
        \includegraphics[scale=0.4, angle=0]{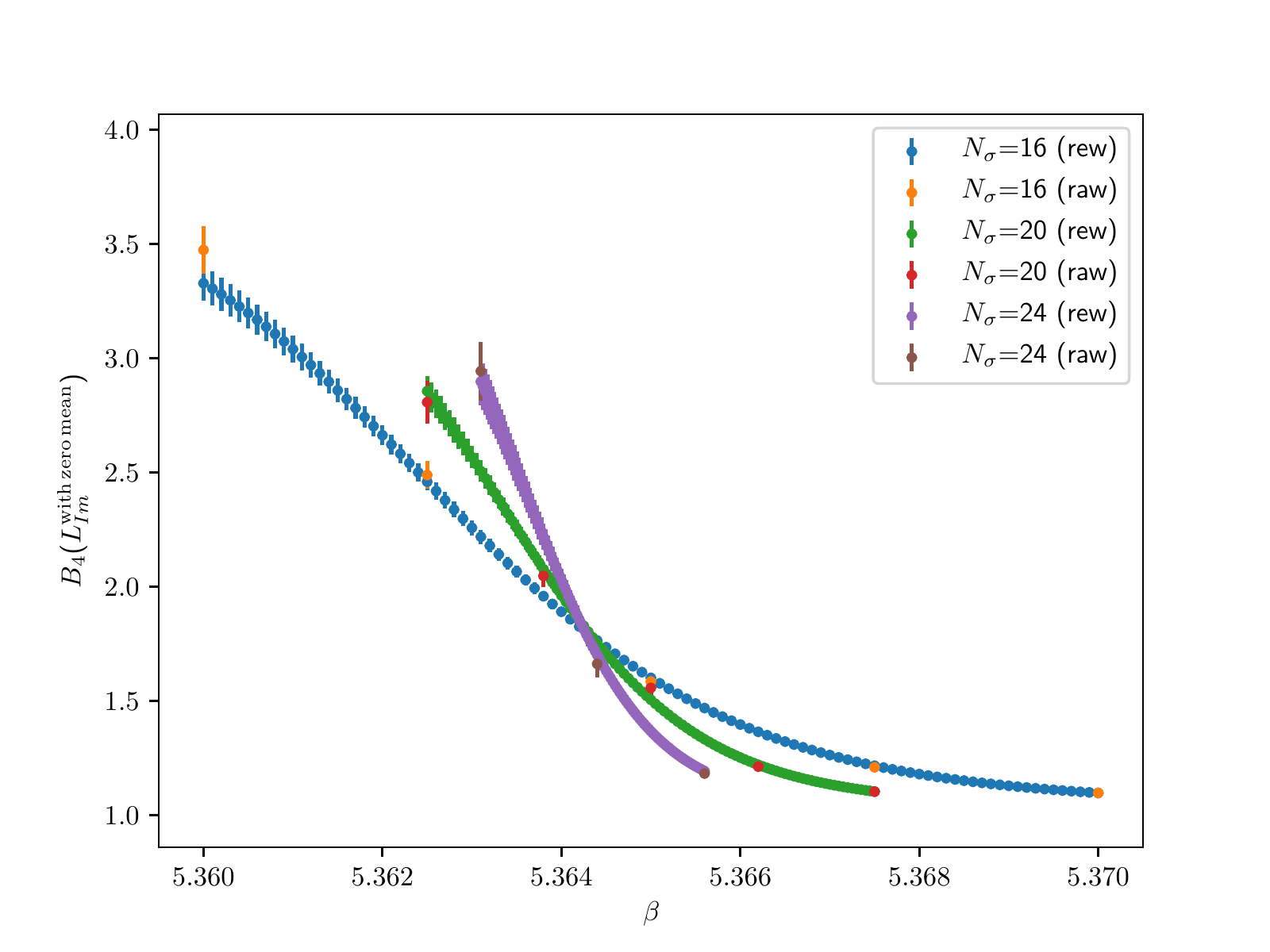}
        \label{fig:scaling_plot1}
        \end{subfigure}
        \begin{subfigure}{.45\textwidth}
  \centering
  \includegraphics[scale=0.4, angle=0]{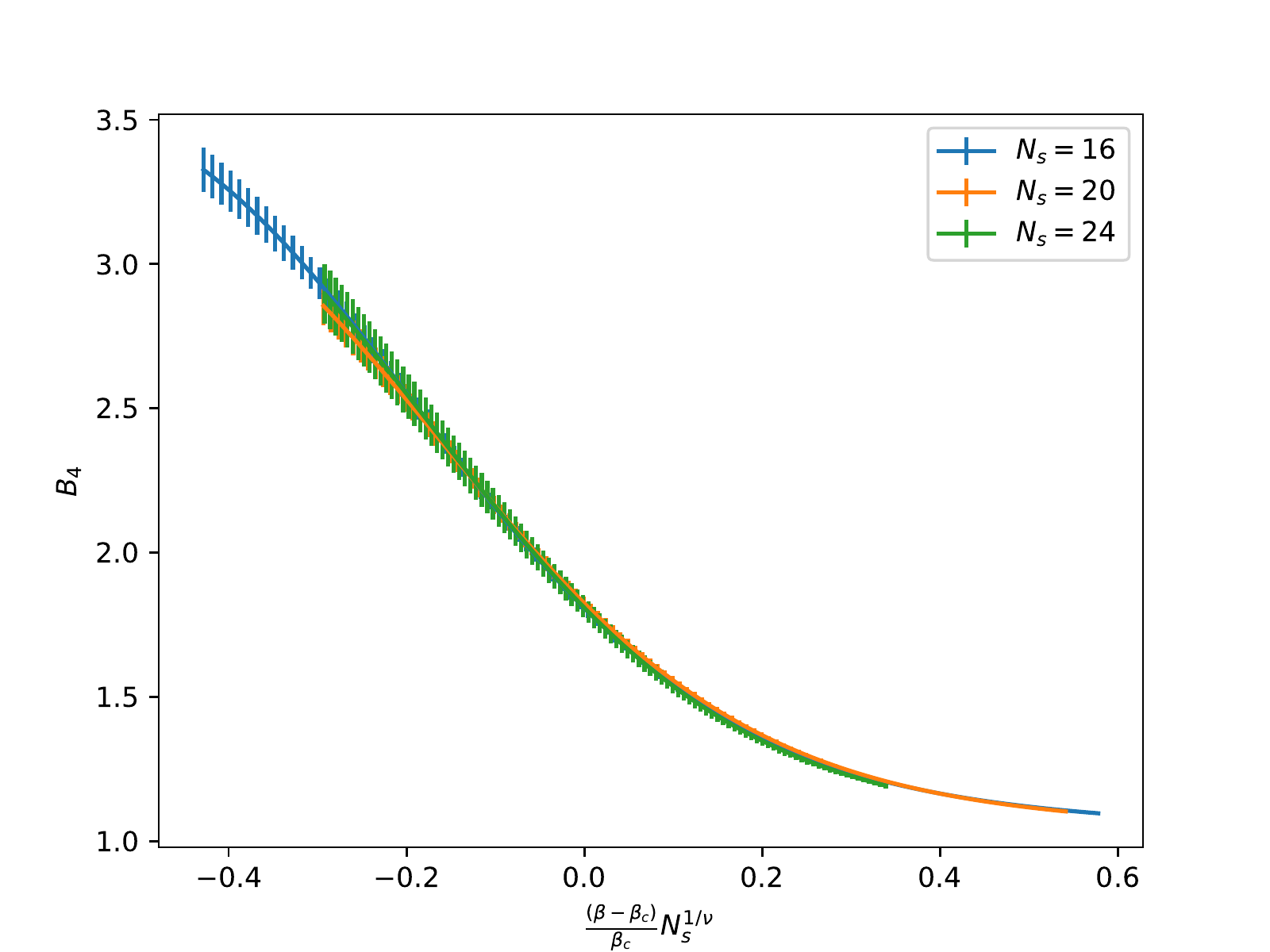}
  \end{subfigure}
  \caption{(Left) The kurtosis as a function of $\beta$ for $N_s=16,20,24$ at $am=0.04$. The points labeled ``raw'' are the simulated points, while the points labeled ``rew'' are the reweighted points obtained using the multi-histogram method \cite{FerrenbergSwendsenMH}. (Right) The collapse of the kurtosis for three different spatial volumes for $am=0.04$ after determining the critical parameters $\beta_c$ and $\nu$ via (\ref{eq:QualityofCollapse}).}
        \label{fig:scaling_plot2}
\end{figure}
\begin{figure}
\centering
\begin{subfigure}{.5\textwidth}
  \centering
  \includegraphics[scale=0.45]{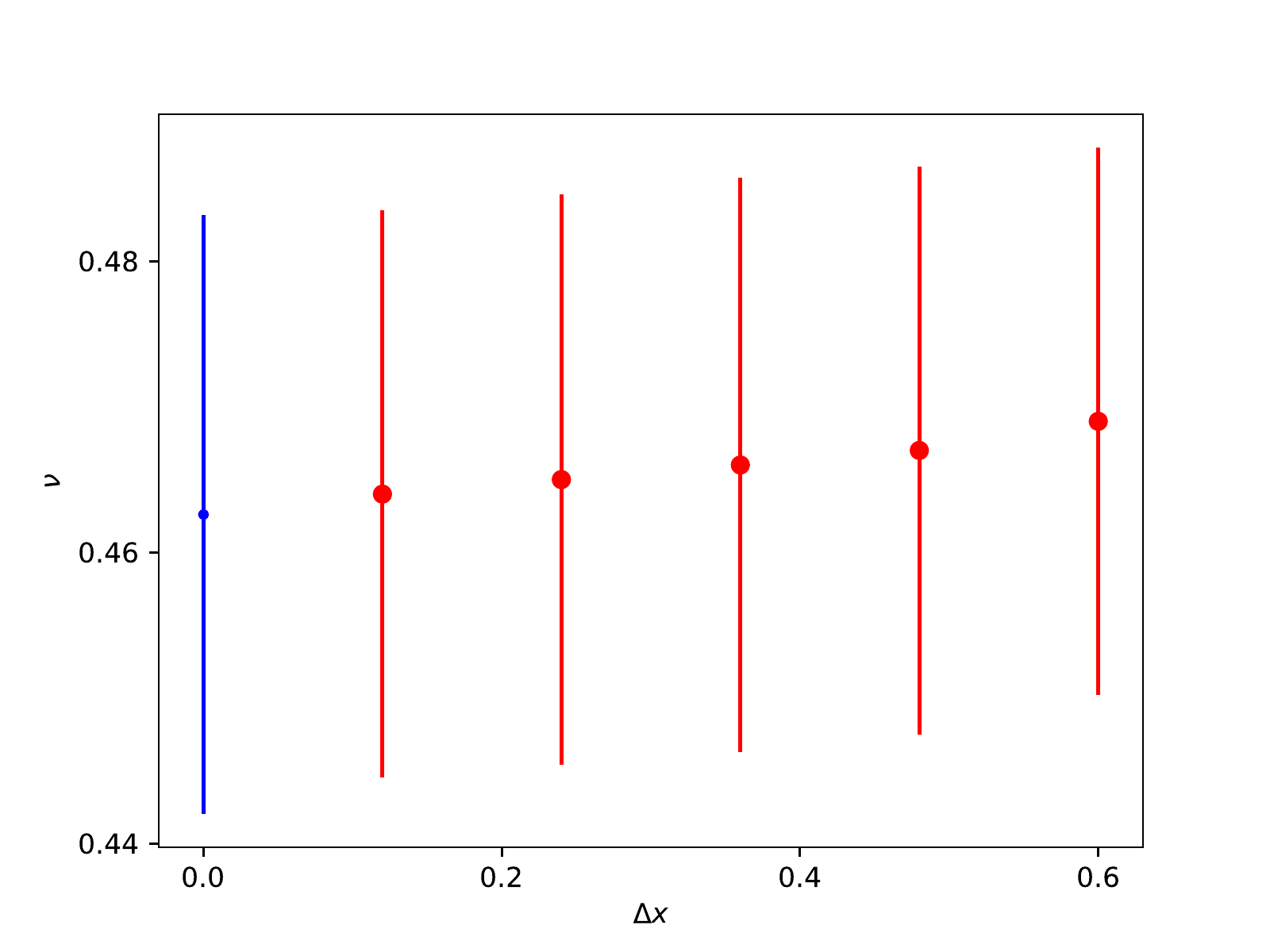}
  \caption{The optimal $\nu$ vs $\Delta x$.}
  \label{fig:sub1}
\end{subfigure}%
\begin{subfigure}{.5\textwidth}
  \centering
  \includegraphics[scale=0.45]{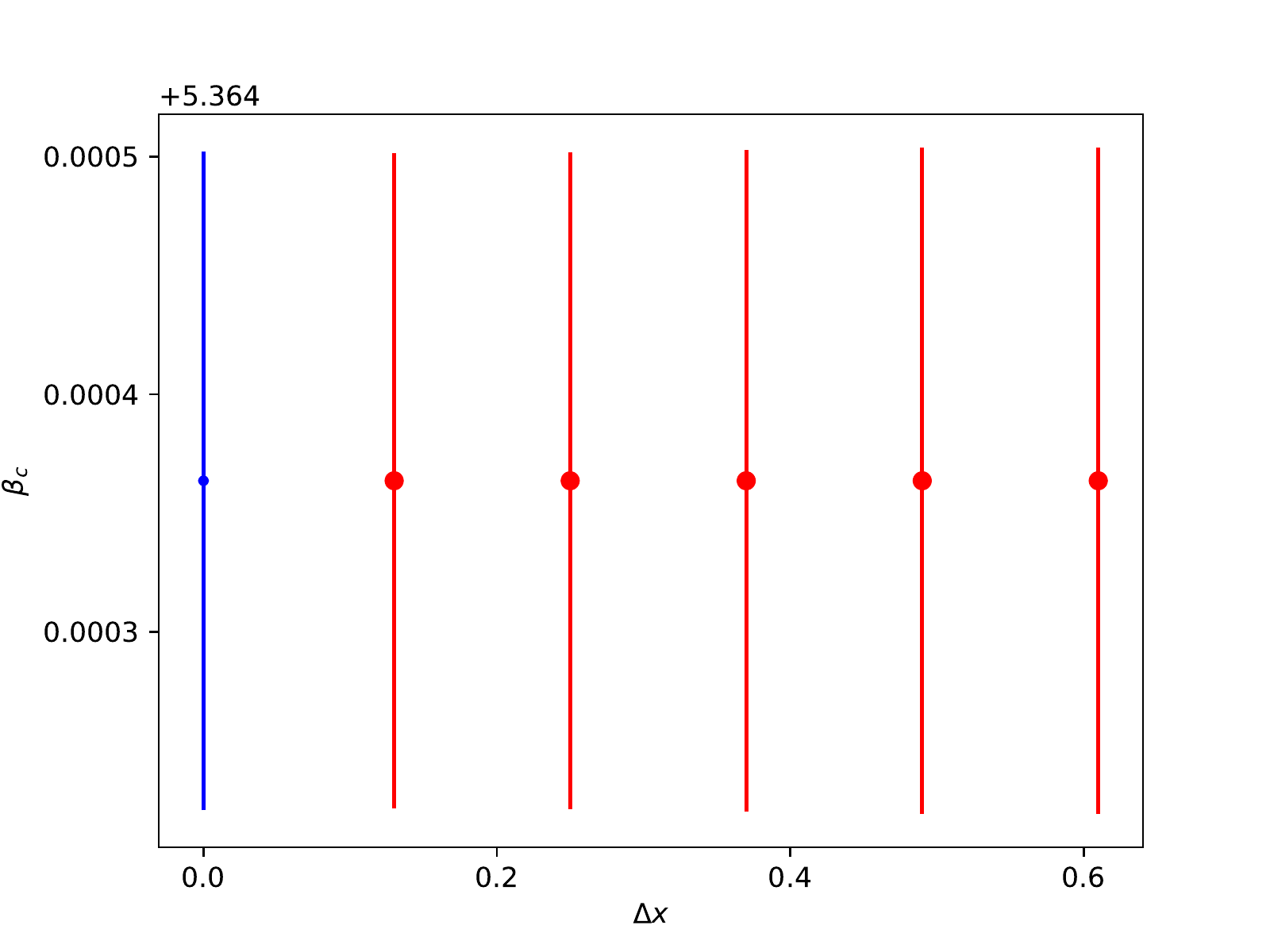}
  \caption{The optimal $\beta_c$ vs $\Delta x$.}
  \label{fig:sub2}
\end{subfigure}
\caption{The dependence of the critical parameters, obtained from the minimization of $Q$, as a function of $\Delta x$ for $am=0.04$ using $N_s=16,20$. The errors on each each point were estimated using $500$ bootstrap estimators for the kurtosis. The result is clearly stable with respect to $\Delta x$, and thus one can safely extrapolate to $\Delta x=0$.}
\label{fig:critical_extrap}
\end{figure}
\begin{figure}
        \centering
        \includegraphics[scale=0.5, angle=0]{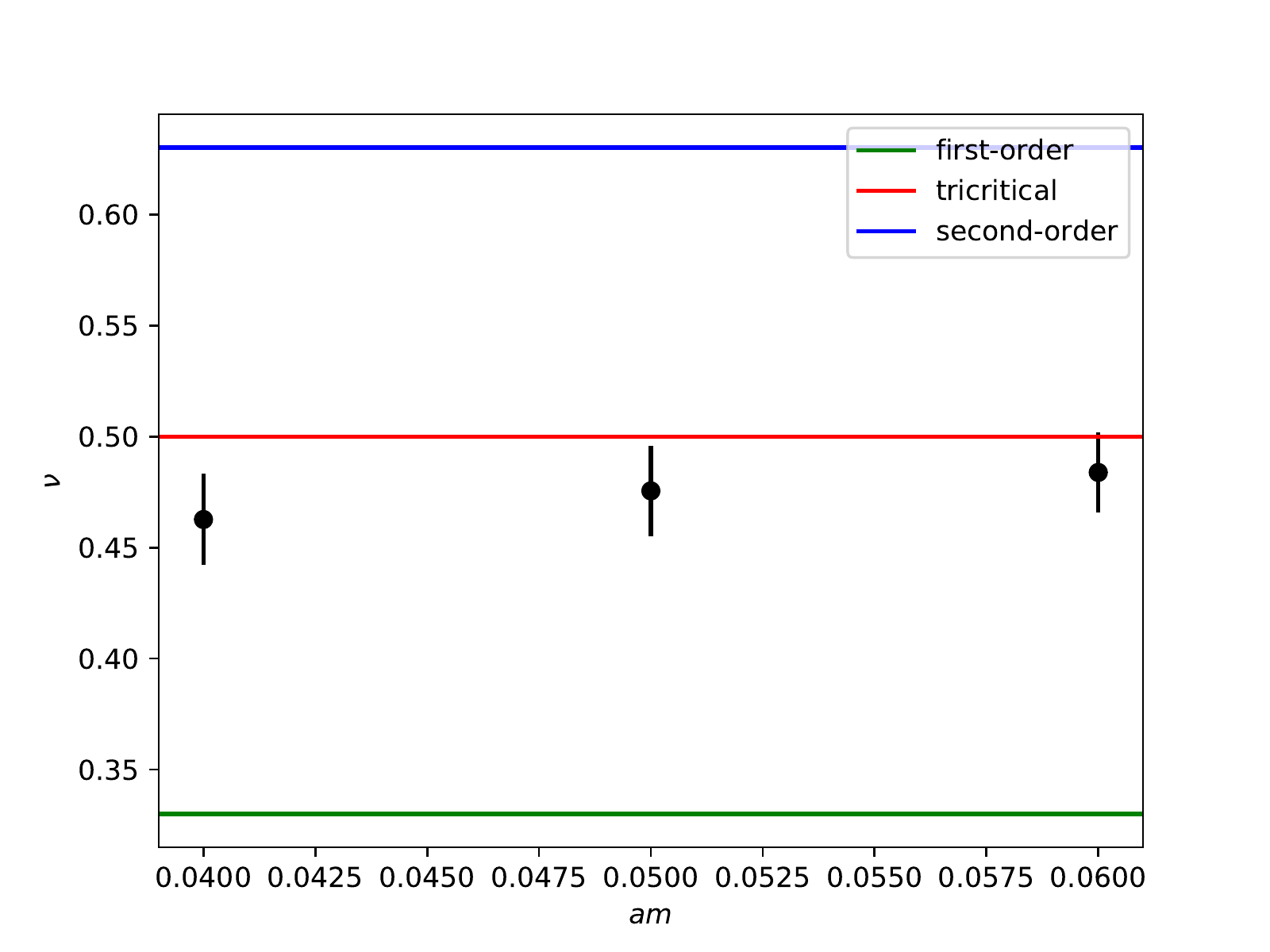}
        \caption{The critical exponent $\nu$ as a function of the bare quark mass $am$. These were results obtained from the quantitative collapse procedure using $N_s=16,20$. One can see that at $am=0.06$, the value of $\nu$ is consistent with the tricritical value.}
        \label{fig:final_nu_plot}
\end{figure}

For each value of the quark mass, a scan in $\beta$ around $\beta_c$ is made. To determine the value of $\beta_c$ as well as the critical exponent $\nu$, we measured the value of the normalized fourth moment of the imaginary part of the volume-averaged Polyakov loop
\beq
B_4 \equiv \frac{\langle \left( \delta \mathcal{O}\right)^4\rangle}{\langle \left( \delta \mathcal{O}\right)^2 \rangle^2}, \quad \delta \mathcal O = \Im P - \langle \Im P \rangle,
\eeq
where we know, \textit{a priori}, that at $\theta_{B,c}$, $\langle \Im P \rangle = 0$. This quantity is commonly referred to as the kurtosis and is expected to obey the following scaling relation around $\beta_c$ 
\beq
B_4(\beta; N_s) = g(x) \equiv  g((\beta-\beta_c) N^{1/\nu}_s), 
\eeq 
where $g$ is a universal scaling function and $x \equiv (\beta-\beta_c) N^{1/\nu}_s$ is the corresponding scaling variable. Thus, by plotting the kurtosis measured on different spatial volumes as a function of the scaling parameter, one hopes to obtain a collapse of the data given the correct values of $\beta_c$ and $\nu$. An example of this collapse of the kurtosis is plotted in fig.~(\ref{fig:scaling_plot2}) for $am=0.04$. The values of $\nu$ and $B_4(\beta_c)$ for the various scenarios occurring at the RW critical point are universal quantities whose values are known at $V\to \infty$. To identify the light tricritical mass from our simulations, these quantities must be determined at finite $V$ as a function of the quark mass. Following previous studies \cite{UpdateColumbiaPlot,newman1999monte}, one can define a quantity which determines the quality of the collapse of the kurtosis in the critical region 
\beq \label{eq:QualityofCollapse}
Q\left(\beta, \nu; \{N^{(i)}_s\}\right) = \frac{1}{2\Delta x} \int^{+\Delta x}_{-\Delta x} \left[ N_V \sum_i \left(B_4(x(\beta,\nu,N^{(i)}_s))\right)^2 - \left( \sum_i B_4(x(\beta,\nu,N^{(i)}_s) \right)^2 \right]dx.
\eeq
The quantity $Q$ estimates the average variance of the data and is minimized as a function of $\beta_c$ and $\nu$. The integration interval is chosen to be symmetric and the results at finite $\Delta x$ are extrapolated to zero to give our final estimate. An example of this procedure is shown in fig.~(\ref{fig:critical_extrap}). We note that this minimization procedure, particularly for the first-order region, is very sensitive to finite-volume effects which we are still currently trying to quantify. 
In practice, three different spatial volumes were used in (\ref{eq:QualityofCollapse}) for each value of the quark mass. Gathering these results, we plot our $\nu$, determined by the extrapolated, minimized variance, as a function of bare quark mass in fig.~(\ref{fig:final_nu_plot}). These data suggest that the light tricritical mass increases as we turn on a nonzero imaginary isospin chemical potential. 

\section{Polyakov loop dynamics}
\begin{figure}
        \centering
        \includegraphics[scale=0.5, angle=0]{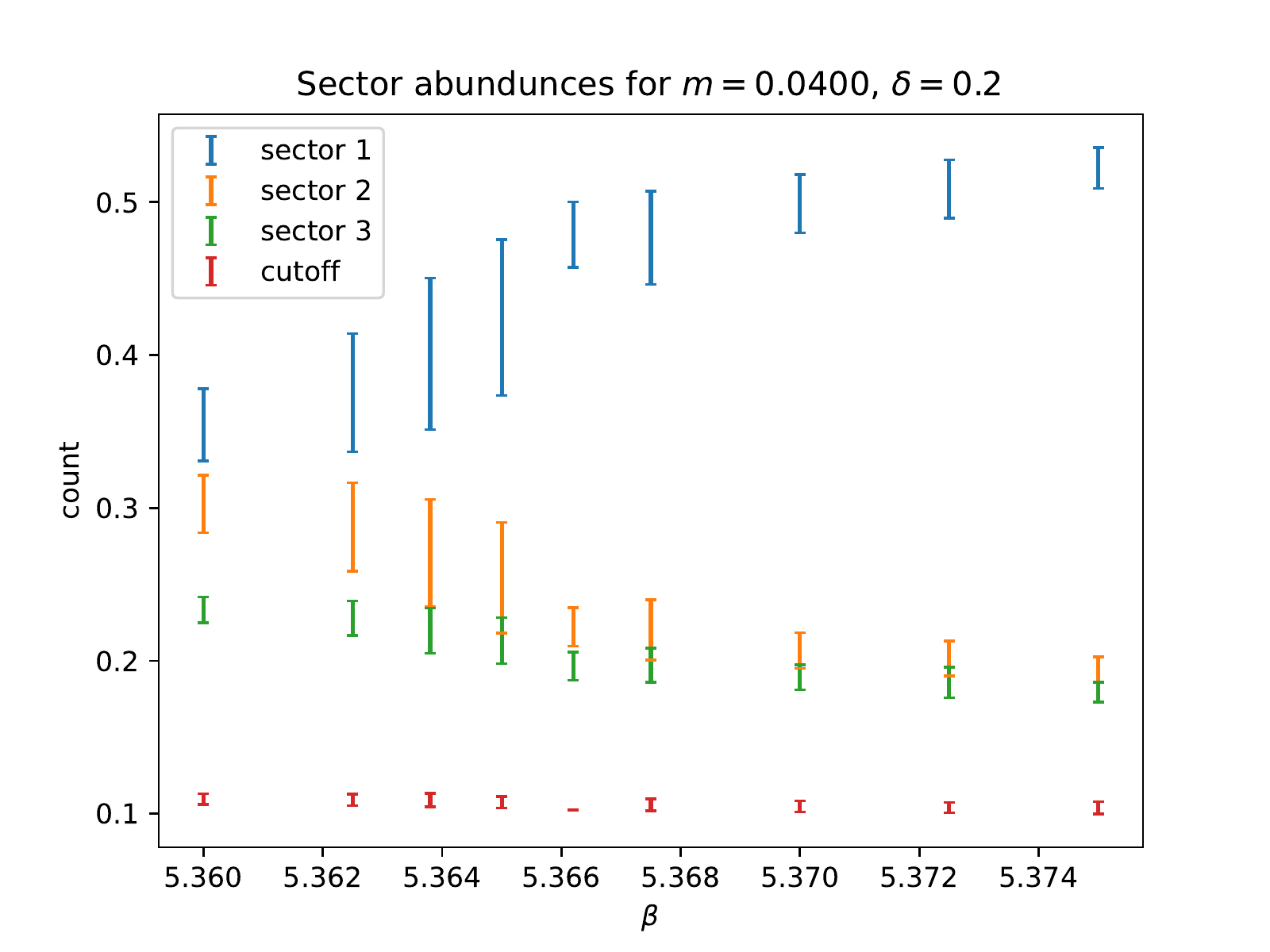}
        \caption{A plot of the relative abundances of the sites corresponding to a given sector $n(x) \in \{1, 2, 3\}$ as a function of $\beta$ for $am=0.04$ for configurations whose spatially-averaged Polyakov loop belong to $n=1$. One notices that near $\beta_c$, the relative abundances of the three sectors, labeled by $n$, begin to diverge from one another.}
        \label{fig:abundances_plot}
\end{figure}

In our simulations, we produce scans in temperature by tuning the parameter $\beta$. As one goes from small $T$ (small $\beta$) where the RW transition is a crossover to large $T$ (large $\beta$) where it is first order, the distribution of $\Im P$ goes from being a Gaussian centered about $0$ to a bimodal distribution whose mean is also zero. During the generation of the gauge ensembles at $\beta>\beta_c$, one samples for long periods of time regions in phase space corresponding to one or the other minimum of the free energy. The probability that a tunneling event occurs between these two regions of phase spaces strongly depends on the volume. In order to understand the dynamics underlying these events, we have studied the clustering properties of the local Polyakov loop, $P(\vec{x})$. Decomposing the Polyakov loop as $P = |P|e^{i\phi}$, we can assign the volume-average as well as each site in $V=N^3_s$ to a sector, $n$, which can be defined according to the following definition 
\beq
n = \begin{cases} 
 3, & \text{$\phi \in  [ -\frac{\pi}{3}+\delta, \frac{\pi}{3}-\delta ]$} \\  
 2 & \text{$\phi \in  [ \frac{\pi}{3}+\delta, \pi-\delta ]$} \\  
 1 & \text{$\phi \in  [ \pi+\delta, \frac{5\pi}{3}-\delta ]$}  
 \end{cases},
\eeq
where $\delta$ is a cutoff parameter which determines the number of spatial sites whose Polyakov loops are discarded in the ensuing analysis.
Following previous studies of pure Yang-Mills theory \cite{Fortunato2003,GATTRINGER2010179,EndrodiGattringer2014} as well as studies with dynamical quarks \cite{Danzer2011}, we have looked at several observables characterizing the local dynamics of the Polyakov loop. In fig.~(\ref{fig:abundances_plot}), we show the abundances of the sectors as a function of $\beta$ on gauge configurations whose spatial average lies in sector 1. The plot illustrates how the dominant sector becomes populated at the expense of the other sectors as we cross the RW endpoint. One can further characterize the sizes of the clusters which are defined by a group of spatial lattice sites each having a neighbor belonging to the same sector $n$. A further investigation of the properties of these clusters is underway.

\section{Summary and Outlook}

We have studied the nature of the RW endpoint at nonzero imaginary isospin and baryon chemical potential in light of its indirect implications for the chiral region of the QCD phase diagram. Our results indicate that the introduction of a nonzero imaginary isospin chemical potential increases the lower tricritical mass. This might have implications on the computational cost of performing chiral extrapolations. In addition, the clustering of local Polyakov loops was studied as $T$ was varied across the RW endpoint. This provided insight into the gauge field dynamics at $\tilde{\mu}_{B,c}$ and $\tilde{\mu}_I \neq 0$. Note that ours is an exploratory study performed at fixed $N_{\tau}=4$. A systematic analysis of cutoff effects is left for the future. 
\acknowledgments
This work was supported by the Deutsche Forschungsgemeinschaft (DFG, German Research Foundation) – project number 315477589 – TRR 211. The authors acknowledge the use of the Goethe-HLR cluster and also thank the computing staff for their support. The authors also thank A. Sciarra for fruitful discusssions. 
\bibliographystyle{JHEP}
\bibliography{RWrefs}

\end{document}